\documentclass[twocolumn]{aastex631} 

\begin{document}

\title{Dominant Role of Coplanar Inflows in Driving Disk Evolution Revealed by Gas-Phase Metallicity Gradients}

\correspondingauthor{Cheqiu Lyu, Enci Wang}
\email{lyucq@ustc.edu.cn, ecwang16@ustc.edu.cn}

\author[0009-0000-7307-6362]{Cheqiu Lyu}
\affiliation{Department of Astronomy, University of Science and Technology of China, Hefei, Anhui 230026, China}
\affiliation{School of Astronomy and Space Science, University of Science and Technology of China, Hefei, Anhui 230026, China}

\author[0000-0003-1588-9394]{Enci Wang}
\affiliation{Department of Astronomy, University of Science and Technology of China, Hefei, Anhui 230026, China}
\affiliation{School of Astronomy and Space Science, University of Science and Technology of China, Hefei, Anhui 230026, China}

\author[0000-0003-1632-2541]{Hongxin Zhang}
\affiliation{Department of Astronomy, University of Science and Technology of China, Hefei, Anhui 230026, China}
\affiliation{School of Astronomy and Space Science, University of Science and Technology of China, Hefei, Anhui 230026, China}

\author[0000-0003-0939-9671]{Yingjie Peng}
\affiliation{Department of Astronomy, School of Physics, Peking University, 5 Yiheyuan Road, Beijing 100871, China}
\affiliation{Kavli Institute for Astronomy and Astrophysics, Peking University, 5 Yiheyuan Road, Beijing 100871, China}

\author[0000-0002-9373-3865]{Xin Wang}
\affiliation{School of Astronomy and Space Science, University of Chinese Academy of Sciences (UCAS), Beijing 100049, China}
\affiliation{Institute for Frontiers in Astronomy and Astrophysics, Beijing Normal University, Beijing 102206, China}
\affiliation{National Astronomical Observatories, Chinese Academy of Sciences, Beijing 100101, China}

\author[0009-0009-2660-1764]{Haixin Li}
\affiliation{Department of Astronomy, University of Science and Technology of China, Hefei, Anhui 230026, China}
\affiliation{School of Astronomy and Space Science, University of Science and Technology of China, Hefei, Anhui 230026, China}

\author[0009-0006-7343-8013]{Chengyu Ma}
\affiliation{Department of Astronomy, University of Science and Technology of China, Hefei, Anhui 230026, China}
\affiliation{School of Astronomy and Space Science, University of Science and Technology of China, Hefei, Anhui 230026, China}

\author[0009-0008-1319-498X]{Haoran Yu}
\affiliation{Department of Astronomy, University of Science and Technology of China, Hefei, Anhui 230026, China}
\affiliation{School of Astronomy and Space Science, University of Science and Technology of China, Hefei, Anhui 230026, China}

\author[0009-0004-5989-6005]{Zeyu Chen}
\affiliation{Department of Astronomy, University of Science and Technology of China, Hefei, Anhui 230026, China}
\affiliation{School of Astronomy and Space Science, University of Science and Technology of China, Hefei, Anhui 230026, China}

\author[0009-0004-7042-4172]{Cheng Jia}
\affiliation{Department of Astronomy, University of Science and Technology of China, Hefei, Anhui 230026, China}
\affiliation{School of Astronomy and Space Science, University of Science and Technology of China, Hefei, Anhui 230026, China}

\author[0000-0002-7660-2273]{Xu Kong}
\affiliation{Department of Astronomy, University of Science and Technology of China, Hefei, Anhui 230026, China}
\affiliation{School of Astronomy and Space Science, University of Science and Technology of China, Hefei, Anhui 230026, China}


\begin{abstract}

Using spatially resolved spectroscopic data from the MaNGA sample, we investigate the parameters influencing the radial gradients of gas-phase metallicity ($\nabla\log(\mathrm{O/H})$), to determine whether disk formation is primarily driven by coplanar gas inflow or by the independent evolution of distinct regions within the disk. Our results show that $\nabla \log(\mathrm{O/H})$ strongly correlates with local gas-phase metallicity at a given stellar mass, with steeper gradients observed in metal-poorer disks. This trend supports the coplanar gas inflow scenario, wherein the gas is progressively enriched by in situ star formation as it flows inward. In contrast, the radial gradient of stellar mass surface density shows very weak correlations with $\nabla \log(\mathrm{O/H})$, which is inconsistent with the independent evolution mode, where gas inflow, star formation, and metal enrichment occur independently within each annulus of the disk. Furthermore, we find that $\nabla \log(\mathrm{O/H})$ is also closely correlated with an indicator of local gas turbulence $\sigma_{\mathrm{gas}}/R_{\mathrm{e}}$, highlighting the competing roles of turbulence and coplanar inflow in shaping metallicity gradients. Our results provide indirect observational evidence supporting coplanar gas inflow as the driving mechanism for disk evolution.

\end{abstract}

\keywords{disk formation --- gas inflow --- gas-phase metallicity --- star-forming galaxy}

\section{Introduction} \label{sec:intro}

The formation and evolution of disks is one of the hot-debated topics in galaxy evolution. Star-forming galaxies exhaust their cold gas supply on relatively short timescales of a few billion years or less, necessitating continuous gas accretion to sustain star formation \citep[e.g.,][]{Bigiel2011, Fraternali2012, Wang2013, Madau2015}. Hydrodynamical simulations generally suggested that gas inflow onto galaxies occurs in a coplanar and broadly corotating manner with the gas disk, while outflows predominantly leave the disk perpendicular to its plane \citep{Keres2005, Stewart2011, Danovich2015, Stewart2017, Peroux2020, Stern2020, Trapp2022, Hafen2022, Gurvich2023}. In contrast, some analytical models based on the bathtub model formalism—where the gas in a galaxy is treated as a reservoir, with the star formation rate regulated by the rates of gas inflow and outflow \citep[e.g.,][]{Bouche2010, Dave2012, Lilly2013}—assumed that all annuli within the disk evolve independently, without significant radial inflows of cold gas along the disk \citep[e.g.,][]{Chiappini2001, Schonrich2017, Lian2018, Belfiore2019b}. However, observational evidence for separating these two disk formation modes is still lacking up to now.  
Therefore, distinguishing between these two modes of evolution of disks, namely coplanar gas inflow or annuli that evolve independently, is critical to understanding the mechanisms driving disk formation and growth.

Despite its importance, observational evidence for significant gas inflows in disks remains sparse. A key challenge is that gas inflows in disks exhibit fairly low radial velocities, typically a few km/s \citep{Warner1973, Trachternach2008, Speights2019, Bisaria2022}. For instance, \citet{Schmidt2016}, analyzing ten galaxies from the THINGS survey \citep[The H I Nearby Galaxy Survey;][]{Walter2008}, found minimal or undetectable radial velocities ($\sim$10 km/s or less) in neutral hydrogen, even at the outer edges of disks. Similarly, \citet{DiTeodoro2021} investigated 54 local disk galaxies using 21 cm emission data cubes and found no evidence of systematic radial inflows or increasing radial velocities in the outermost regions. Using a modified accretion disk model, \citet{Wang2023} suggested that coplanar gas inflow velocities within two disk scale lengths are likely only a few km/s, increasing gradually with radius and reaching up to 50–100 km/s in the outermost disk regions. 

Studying the gas-phase metallicity profile is a promising way for uncovering evidence of gas inflows and examining the modes of disk evolution, acting as a powerful chemical diagnostic closely tied to disk formation and gas kinematics.
Cold gas inflows and outflows influence metal enrichment at different galactic radii, resulting in distinct radial gradients of gas-phase metallicity \citep[e.g.,][]{Mayor1981, Lacey1985, Sommer-Larsen1989, Goetz1992, Thon1998, Portinari2000, Spitoni2011, Cavichia2014, Kubryk2015, Pezzulli2016, Sharda2021, Wang2022b, Wang2024, Chen2024}. These gradients are shaped by physical processes that accompany disk formation and evolution, including star formation, gas inflows, gas outflows, mergers, stellar migration, and environmental factors, such as ram pressure or tidal stripping \citep[e.g.,][]{Tinsley1980, Tremonti2004, Finlator2008, Mannucci2010, Rupke2010,Pilkington2012, Jones2013, Peng2014, Ho2015, Tissera2016, Bahe2017, Maiolino2019, Wang2019, Wang2021, Wang2022a, Wang2022b, Sharda2023, Lyu2023, He2024, Cheng2024, Venturi2024}. For instance, the presence of anomalously low-metallicity regions in star-forming galaxies underscores the impact of gas accretion events on local metallicity and disk formation \citep{Hwang2019}. The metallicity gradient thus encodes a wealth of information on these processes, providing critical insights into disk assembly and evolution.

In the local Universe, most galaxies exhibit negative metallicity gradients, with metallicity decreasing from the center to the outskirts. This ``negative metallicity gradient'' \citep[e.g.,][]{Searle1971, Zaritsky1994, vanZee1998, Sanchez2014, Belfiore2017, Kreckel2019, Sanchez-Menguiano2020} aligns with the standard inside-out disk formation scenario \citep[e.g.,][]{Fall1980}, where earlier star formation in the inner regions allows more time for chemical enrichment compared to the outer areas \citep[e.g.,][]{Samland1997, Dave2011, Gibson2013, Hemler2021, Tissera2022, Jia2024}. The steepening of the metallicity gradient can be explained by significant redistribution of metals during galaxy formation, driven by several mechanisms, such as powerful feedback effects, prominent radial flows, or stochastic accretion/dilution processes \citep{Dekel2013, Grisoni2018, Maiolino2019}. In addition, negative gradients naturally arise as in situ star formation enriches gas flowing inward \citep[e.g.,][]{Wang2022b, Wang2024}. However, feedback-driven outflows, turbulence, and other processes can flatten these gradients \citep[e.g.,][]{Gibson2013, Marinacci2014, Wang2022b}. Conversely, metallicity gradients may even invert (become positive) due to processes such as the dilution of enriched gas by pristine intergalactic medium (IGM) accretion, galactic fountains, or mergers \citep[e.g.,][]{Cresci2010, Kewley2010, Rupke2010, Brook2012, Fu2013, Troncoso2014}. As the product of various physical processes, metallicity gradients are complex yet invaluable indicators of galaxy assembly history, gas dynamics, feedback, and accretion.

In this work, we use gas-phase metallicity radial gradients (hereafter ``metallicity gradient'' for short) to investigate whether disk formation is dominated by coplanar gas inflows or by independent evolution of disk annuli. The spatially resolved sample enables us to test these competing scenarios. This paper is organized as follows: Section \ref{sec:data} describes the sample selection and methods for determining galaxy properties. Section \ref{sec:motivation} states the motivation of this work. In Section \ref{sec:results}, we analyze correlations with other galaxy properties. Section \ref{sec:summary} summarizes our findings and discusses their implications. Throughout this work, gas-phase metallicity is expressed as the oxygen abundance, 12 + log(O/H), where O/H is the ratio of oxygen to hydrogen element number densities. We assume a flat cold dark matter cosmology model with $\Omega_m=0.27, \Omega_\Lambda=0.73$, and $h=0.7$ when computing distance-dependent parameters.

\section{Data} \label{sec:data}

\subsection{The MaNGA survey}

Our parent sample is derived from the Sloan Digital Sky Survey IV \citep[SDSS IV;][]{Blanton2017} Mapping Nearby Galaxies at Apache Point Observatory \citep[MaNGA;][]{Bundy2015} survey. MaNGA provides two-dimensional spectra for approximately 10,000 galaxies with redshifts in the range $0.01 < z < 0.15$, utilizing the spectrographs on the 2.5-meter Sloan Telescope at Apache Point Observatory \citep{Gunn2006, Smee2013}. The survey covers wavelengths from 3600 to 10300 \AA\ with a spectral resolution of $R \sim 2000$. Spatial coverage typically extends beyond 1.5 $R_\mathrm{e}$ for individual galaxies, with a spatial resolution of 1–2 kpc. Detailed descriptions of the MaNGA instrument design, testing, and assembly are provided in \citet{Drory2015}.

We adopt the emission line flux and stellar and gas kinematics measurements from the MaNGA Data Analysis Pipeline (DAP) \citep{Westfall2019, Belfiore2019a}. The DAP employs adaptive Voronoi spatial binning \citep{Cappellari2003} to achieve a target signal-to-noise ratio (S/N) of approximately 10 in the stellar continuum. After measuring stellar kinematics, the DAP fits Gaussian profiles to derive emission line fluxes for each spaxel using the full spectral fitting code \texttt{pPXF} \citep{Cappellari2004, Cappellari2017}.

\subsection{Sample selection}  

We cross-match the MaNGA galaxy sample with the GALEX-SDSS-WISE Legacy Catalog \citep[GSWLC;][]{Salim2016, Salim2018}, which provides galaxy stellar masses derived from spectral energy distribution (SED) fitting of ultraviolet, optical, and infrared photometry. Our selection criteria include galaxies with stellar mass and star formation rate (SFR) data available in the GSWLC-M2 catalog. We adopt the {\tt SERSIC\_TH50} as effective radii ($R_\mathrm{e}$) from the NASA-Sloan Atlas \citep[NSA;][]{Blanton2011} and convert it to physical scale assuming a flat cold dark matter cosmology model with $\Omega_{m}, \Omega_{\Lambda},h=0.27,0.73,0.7$. We select galaxies with axis ratios ($b/a$) larger than 0.4, where $b/a$ are taken from {\tt SERSIC\_BA} in NSA catalog. Stellar masses are rescaled to a \citet{Kroupa2001} initial mass function (IMF). We focus on star-forming galaxies, selecting those with $\Delta$MS $>-1$ dex. Here, $\Delta$MS represents the vertical offset from the star-forming main sequence (SFMS):
$$\log(\mathrm{SFR/M_\odot yr^{-1}})=0.71 \times \log(\mathrm{M_\odot}) - 7.27,$$
which is defined in \citet{Ma2024} for MaNGA sample based on an iterative algorithm \citep{Woo2013, Donnari2019, Wang2023}. Specifically, the algorithm begins with an initial guess of a linear function and iteratively selects galaxies within 1 dex of the SFMS. It then refits the slope and intercept until convergence is achieved. This selection results in a final sample of 3,808 galaxies.

The gas velocity dispersion ($\sigma_{\mathrm{gas}}$) is also taken from the MaNGA DAP product. For each galaxy, we select spaxels with good data quality (indicated by {\tt QUALDATA} in MaNGA DAP) and valid H$\alpha$ fluxes within galactocentric radii of 0.5–1.5 $R_\mathrm{e}$. After correcting for instrumental dispersion, we use the median H$\alpha$ velocity dispersion in annuli 0.5–1.5 $R_\mathrm{e}$ for each galaxy, as an indicator of gas turbulence.

\subsection{Radial gradient of gas-phase metallicity}

For each galaxy, we identify star-forming and composite spaxels within galactocentric radii of 0.5–1.5 $R_\mathrm{e}$ using the BPT-NII diagnostic diagram \citep{Baldwin1981, Kewley2001, Kauffmann2003}. Selected spaxels meet the criteria S/N $> 3$ for the fluxes of H$\alpha$, H$\beta$, [OIII]$\lambda$5007, [NII]$\lambda$6583, and [OII]$\lambda$3737, 3729.  
Emission line fluxes are corrected for interstellar reddening using the \citet{Cardelli1989} reddening law with $R_V = 3.1$. Extinction is calculated by comparing the measured H$\alpha$/H$\beta$ flux ratio to the theoretical value of 2.86 \citep{Osterbrock2006}, assuming case B recombination. Spaxels classified as Seyfert or LINERs are excluded. For star-forming and composite spaxels, we determine the H$\alpha$-flux-weighted gas-phase metallicity (12+log(O/H)) using the \texttt{N2S2H$\alpha$} calibrator proposed by \citet{Dopita2016}, which is insensitive to the reddening and ionization parameter via the [NII]/H$\alpha$ term \citep[e.g.,][]{Zhang2017, Hwang2019}. In addition, \citet{Easeman2024} proposed that the \texttt{N2S2H$\alpha$} is preferred when studying the distribution of metals within galaxies because of the near-linear relation with $T_\mathrm{e}$-based measurement. Alternative calibrations, such as S-calibration \citep[\texttt{Scal}, ][]{Pilyugin2016} and \texttt{N2O2} \citep{Dopita2013, Zhang2017} also confirm the trends in this work.

We focus on disk-dominated regions. For each galaxy, the median gas-phase metallicity is calculated by considering all spaxels within each annulus, ranging from 0.5 $R_\mathrm{e}$ to 1.5 $R_\mathrm{e}$ with each annulus having a width of 0.2 $R_\mathrm{e}$, requiring at least 70\% of spaxels to be available within each annulus. Gas-phase metallicity radial gradients, expressed in dex/$R_\mathrm{e}$ ($\nabla \log(\mathrm{O/H})$), are derived using linear fits to the radial profiles in this range. This approach minimizes the impact of smearing effects and avoids significant deviations from linear fits \citep{Sanchez2014, Sanchez-Menguiano2016, Belfiore2017, Boardman2022, Barrera-Ballesteros2023}. Metallicity diagnostics are typically calibrated based on HII region models. However, one may be concerned that the spaxel sample selection criteria (e.g., BPT-NII) could result in non-negligible contamination from diffuse ionized gas (DIG) when measuring chemical abundances from emission line ratios \citep[e.g.,][]{Belfiore2017, ValeAsari2019}. To address this issue, we have tested the requirement that spaxels must have a minimum equivalent width of H$\alpha$ (EW(H$\alpha$)) of 10 \AA\ \citep[e.g.,][]{ValeAsari2019} to derive the metallicity profiles and gradients of each galaxy. Combined with our sample selection criteria, this EW(H$\alpha$) cut removes approximately 24\% of the used spaxels. However, since we calculate the metallicity gradient using the median metallicity value of each annulus and subsequently apply linear fitting, this cut has a negligible impact on our final valid galaxy sample (it only reduces the sample by $\sim$8.1\%, from 3,808 to 3,499 valid galaxies). Furthermore, the impact of this cut on our results is minimal.

\subsection{Stellar mass surface density}

We utilize two-dimensional stellar mass maps from \citet{Lu2023}, derived using \texttt{pPXF}. Stellar population synthesis was performed using the FSPS model \citep{Conroy2009, Conroy2010}, encompassing 43 age and 9 metallicity grids. The datacubes are Voronoi binned to achieve S/N $\sim 30$ prior to spectral fitting. For a given Voronoi bin, spaxels share the same stellar mass.

Stellar mass surface density ($\Sigma_*$) is calculated for each spaxel by dividing the stellar mass by the physical area of the pixel, corrected for inclination effect. To avoid biases from incomplete coverage, the median $\Sigma_*$ is calculated by considering all spaxels within each annulus, ranging from 0.5 $R_\mathrm{e}$ to 1.5 $R_\mathrm{e}$ with each annulus having a width of 0.2 $R_\mathrm{e}$. While the annuli at least 70\% spaxels available are finally adopted to build the radial profile. The radial gradients of $\Sigma_*$, in units of dex/$R_\mathrm{e}$ ($\nabla \log\Sigma_*$), are determined by linear fits to the radial profiles in the same range.

\section{Motivation} \label{sec:motivation}

\begin{figure*}[htb!]
    \centering
    \includegraphics[width=1\linewidth]{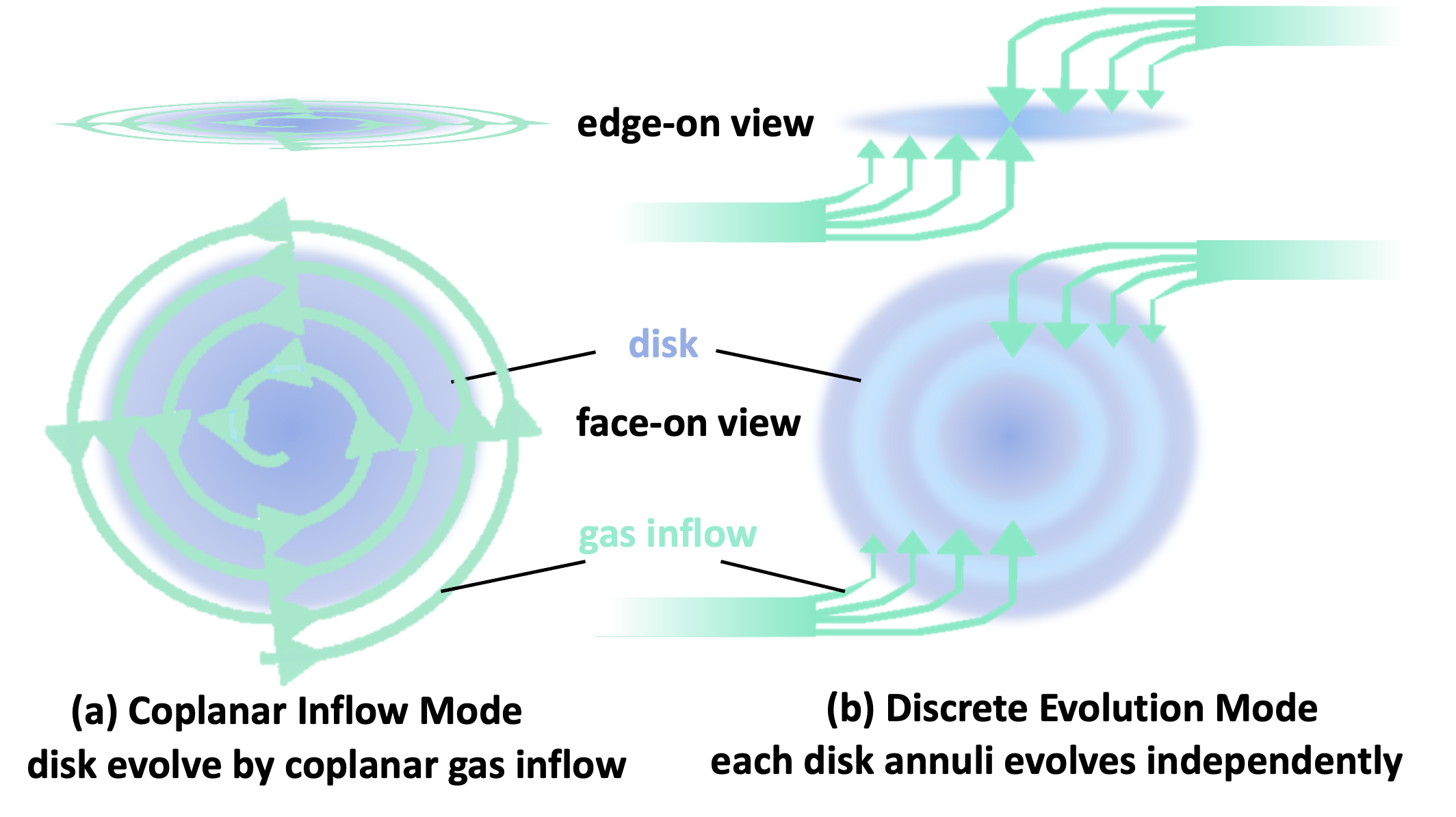}
    \caption{An illustration of the simplified model of two primary modes of galaxy disk formation: \textbf{\tt Coplanar Inflow Mode} (left), where coplanar inflow drives disk evolution and \textbf{\tt Discrete Evolution Mode} (right), where different parts or radii of galaxies evolves independently. The illustration includes both edge-on view (top) and face-on view (bottom) of the gas disk, with the disk planes rendered in blue and the gas inflow in green. Arrows indicate the trajectories of gas inflows.}
    \label{Twomode}
\end{figure*}

The formation of galaxy disks is a complex process. Figure \ref{Twomode} illustrates the simplified model of two primary modes of galaxy disk formation: coplanar inflow driving disk evolution (\textbf{\tt Coplanar Inflow Mode}) and different parts or radii of galaxies evolving independently (\textbf{\tt Discrete Evolution Mode}). These two modes provide distinct frameworks for understanding how gas accretion and internal dynamics influence the evolution of galaxy disks, leaving unique imprints on their metallicity gradients.

(1) \textbf{\tt Coplanar Inflow Mode} posits that galaxy disks evolve through the continuous accretion of gas that flows coplanar and inspiral into the galaxy. The potential mechanism driving coplanar gas inflow is discussed in \cite{Wang2022c}, suggesting that magnetic stresses are the most likely source of the viscosity required to form an exponential star-forming disk.
In this mode, the disk of different radii undergoes a continuous gas accretion process, which sustains star formation and promotes ongoing disk growth. The metallicity gradients are therefore strongly determined by this disk formation mode.  

(2) \textbf{\tt Discrete Evolution Mode} emphasizes the independent evolution of individual annular structures within the galaxy disk. Instead of a continuous gas inflow, the disk is assembled through a series of discrete annuli. Within each annulus, gas accretion (irrespective of origin) drives local star formation and local metal enrichment, resulting in independent evolutionary pathways for each annular structure.

In the {\tt Coplanar Inflow Mode}, the modified accretion disk framework of \citet{Wang2022b} predicts that a negative metallicity gradient naturally arises if radial gas inflow dominates within the disk. The inflowing gas is progressively enriched by in situ star formation as it spirals inward toward the galactic center. In this mode, the metallicity gradient is highly sensitive to the metallicity of the original inflowing gas. Higher metallicity results in less efficient enrichment in log(O/H), leading to a more flattened metallicity gradient.

In contrast, the {\tt Discrete Evolution Mode} suggests a strong correlation between local gas-phase metallicity and local $\Sigma_*$ at different radii, as described by the resolved mass-metallicity relation (rMZR) \citep[e.g.,][]{Rosales-Ortega2012, Sanchez2013, Barrera-Ballesteros2016, Gao2018}. In this mode, the gas-phase metallicity radial gradient is expected to be primarily tied to the gradient of $\Sigma_*$. This could be due to the fact that different $\Sigma_*$ may lead to different mass-loading factors, which is responsible for the metallicity gradients. Therefore, this distinction provides a testable prediction: by examining the gas-phase metallicity radial gradient, we can infer which mode is dominant in driving disk formation.

\section{Results and discussion} \label{sec:results}

\subsection{Radial gradient of gas-phase metallicity}

\begin{figure*}[htb!]
    \centering
    \includegraphics[width=0.49\linewidth]{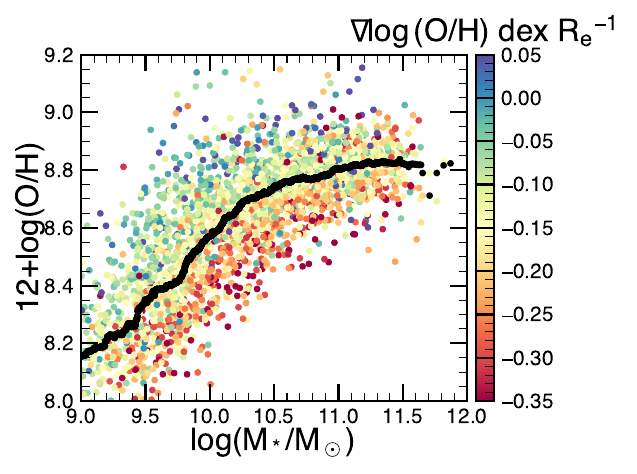}
    \includegraphics[width=0.49\linewidth]{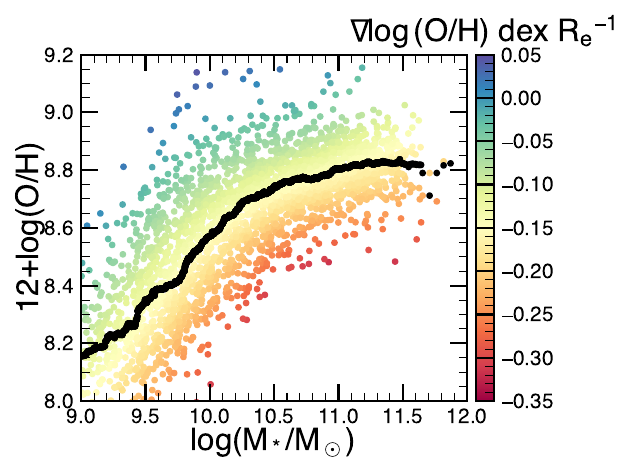}
    \includegraphics[width=1.0\linewidth]{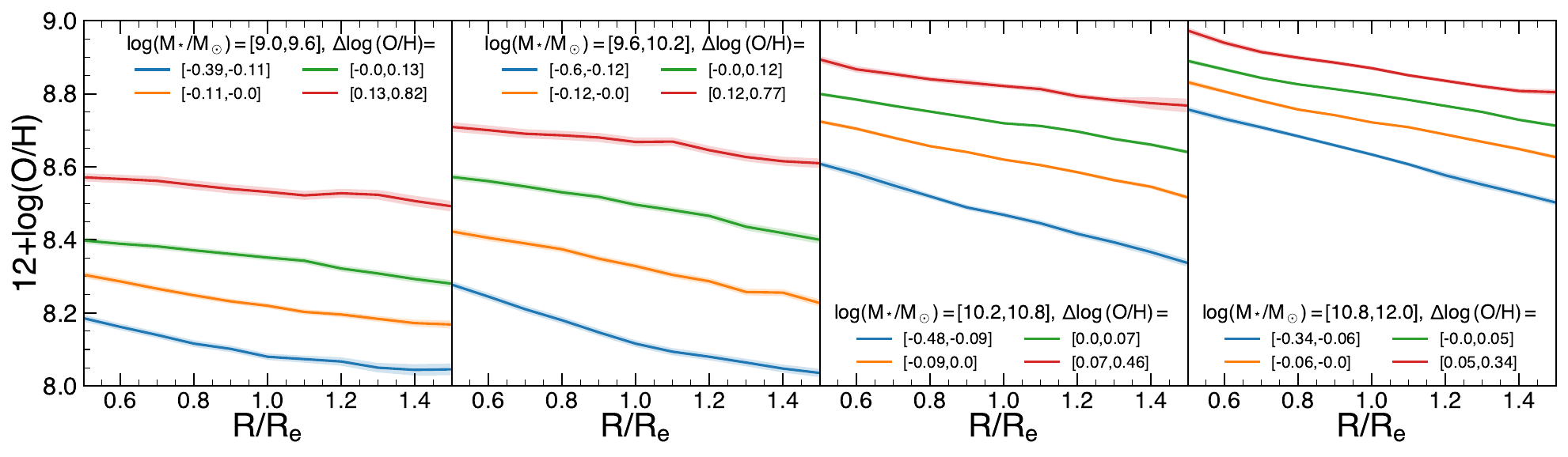}
    \caption{\textbf{Top panels}: Metallicity gradient on the disk as functions of stellar mass and gas-phase metallicity at effective radius. Each dot represents an individual galaxy in our sample, with a redder/bluer color corresponding to a steeper negative/flatter or positive metallicity gradient. The black curve and dots indicate the median y-values in each x-bin with a width of 0.2 dex for each galaxy. The top left panel shows the scatter plot and the top right panel shows the corresponding smooth results using the locally weighted regression method {\tt LOESS} \citep{Cleveland1988, Cappellari2013} to reveal the underlying trend. \textbf{Bottom panels}: Gas-phase metallicity radial profiles in $0.5-1.5$ $R_\mathrm{e}$ considered in this work, subdivided into four stellar mass bins (from left to right: $\log(M_*/\mathrm{M_\odot})=9.0-9.6,9.6-10.2,10.2-10.8,10.8-12.0$). For each bin, the blue, orange, green, and red curves correspond to median profiles of galaxies in four quantiles of metallicity excess ($\Delta \log \mathrm{(O/H)}$). The shaded areas indicate the uncertainties of the median values.}
    \label{MZR}
\end{figure*}

To explore the potential impact of gas coplanar inflow on metallicity distribution, we examine the gas-phase metallicity gradient $\nabla \log(\mathrm{O/H})$ as a function of stellar mass and local gas-phase metallicity at the effective radius, in the top panels of Figure \ref{MZR}. Overall, the local gas-phase metallicity at the effective radius increases with stellar mass, reflecting a local manifestation of the well-established mass-metallicity relation \citep[MZR, e.g.,][]{Tremonti2004, Mannucci2010, Steidel2014, Wuyts2014, Sanchez2019, Ma2024}. At a given stellar mass, galaxies with lower gas-phase metallicity tend to exhibit more negative or steeper $\nabla \log(\mathrm{O/H})$. This trend aligns with the predictions of the modified accretion disk model proposed by \citet{Wang2022b}, wherein inflowing gas is progressively enriched by in situ star formation as it slowly spirals inward toward the disk center. Consequently, higher metallicity in the disk reduces the difference between the metallicities of the outer and inner regions, leading to a relatively flatter $\nabla \log(\mathrm{O/H})$. \citet{Jara-Ferreira2024} found that low-mass galaxies with strong positive metallicity gradients are less enriched than the median mass-metallicity relation, while those with strong negative gradients are more enriched. This connection between metallicity and metallicity gradient is observed in both the MaNGA sample and the \texttt{EAGLE} simulations, and is consistent with our results. Furthermore, this mass-metallicity-gradient relation also follows the mass-size-metallicity relationship \citep[e.g.,][]{Ellison2008, D'Eugenio2018} combined with the mass-size-gradient trend reported in \citet{Boardman2021}, where at a given stellar mass, larger galaxies on average possess lower gas-phase metallicities and display steeper metallicity gradient than smaller galaxies. The relation between stellar mass, gas-phase metallicity, and metallicity gradient for MaNGA data has been also presented by \citet{Franchetto2021}. They found that, at a given stellar mass, metallicity profiles tend to flatten as metallicity increases, which is consistent with the trends shown in Figure \ref{MZR}. Additionally, they observed that the increasing trend of metallicity is also found to be similar to the decreasing trend of gas fraction on the stellar mass–metallicity gradient plane.

To account for the mass dependence of the metallicity gradient, we calculate the metallicity excess ($\Delta \log(\mathrm{O/H})$) for each galaxy, defined as the deviation of the metallicity from the median value for galaxies of the same stellar mass (represented by the black curve and dots in the top panels of Figure \ref{MZR}). By removing the mass dependence, $\Delta \log(\mathrm{O/H})$ provides a more direct measure of the relative enrichment of the interstellar medium (ISM) at the effective radius of each galaxy. Although its value can be influenced by calibration methods and measurement indicators, the metallicity excess reflects the balance of metal production and transport within star-forming galaxy disks. As such, it is a fundamental parameter for understanding galactic chemical evolution on the disk.

To intuitively illustrate the relationship between metallicity excess and the radial metallicity profiles of galaxies, we present the median metallicity radial profiles for the $R_\mathrm{e}$ range in the bottom panels of Figure \ref{MZR}. These profiles are shown across different stellar mass bins (individual panels) and metallicity excess bins (distinguished by colors). From left to right, as stellar mass increases, the overall normalization of the metallicity profiles rises systematically. Within each stellar mass bin, the profiles become progressively flatter as the metallicity excess increases (from blue to red curves). This trend demonstrates the strong correlation between local metallicity excess and the metallicity gradient in the disk.
\begin{figure*}[htb!]
    \centering
    \includegraphics[width=0.49\linewidth]{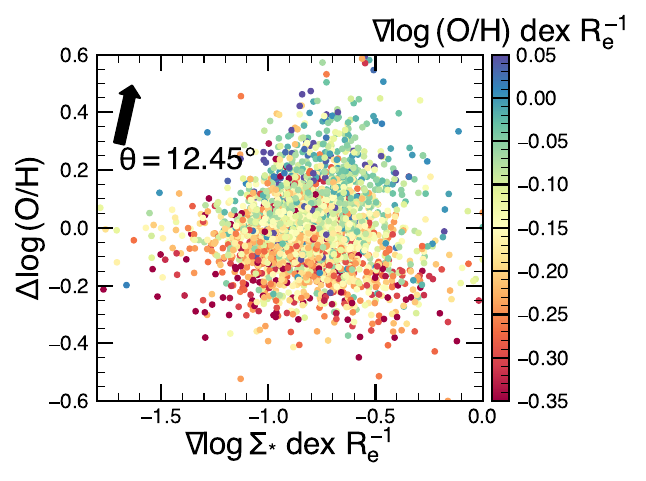}
    \includegraphics[width=0.49\linewidth]{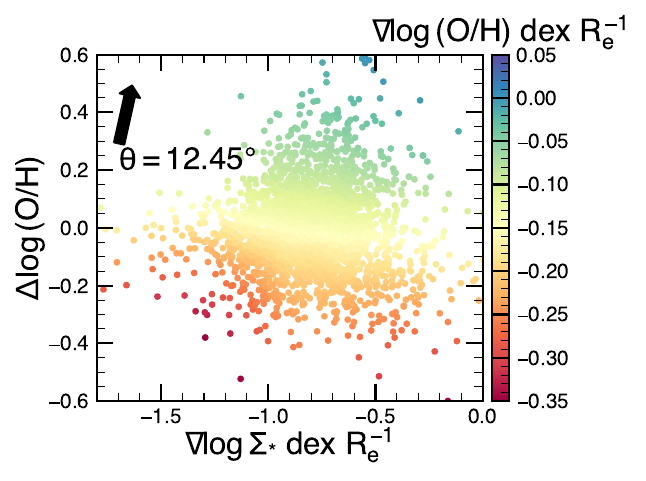}
    \caption{Metallicity gradient as functions of metallicity excess at the effective radius and gradient of $\Sigma_*$. The left and right panels show the scatter plot and corresponding smooth results, respectively. The dots and colors are similar to the top panels of Figure \ref{MZR}. The color gradients, indicated by arrows, represent the directions of increasing $\nabla \log(\mathrm{O/H})$.}
    \label{Zgasexc_SDslope_ccZgasslope}
\end{figure*}

Figure \ref{Zgasexc_SDslope_ccZgasslope} examines the metallicity gradients
as functions of both $\Delta \log(\mathrm{O/H})$ and the gradient of $\Sigma_*$. The results show that $\nabla \log(\mathrm{O/H})$ is only weakly correlated with $\nabla\log \Sigma_*$, exhibiting a Pearson correlation coefficient of $\sim$0.058 (with $p$-value of 0.0011), compared to its much stronger correlation with $\Delta \log(\mathrm{O/H})$. This distinction is further supported by the angles of partial correlation \citep[e.g.,][]{Bait2017,Bluck2019,Baker2023}, as indicated by the black arrows. Furthermore, the weak correlation between $\Delta \log(\mathrm{O/H})$ and $\nabla\log \Sigma_*$ shows that the two quantities are not degenerate. Therefore, the metallicity gradient is not closely linked to the stellar mass surface density gradient, effectively ruling out the dominant role of {\tt Discrete Evolution Mode} in driving disk formation.

\subsection{Local gas turbulence}

The gas disk can be highly turbulent due to stellar winds and supernova explosions, which contribute to radial gas mixing and reduce metallicity gradients. Processes such as turbulence, stellar feedback, accretion, mergers, and satellite interactions mix metals in the ISM, generally flattening metallicity gradients. This effect is particularly pronounced at early times when turbulence is higher and the ISM is geometrically thicker \citep{Bird2012, Bird2021, McCluskey2024, Graf2024}. Additional detailed mechanisms that influence metallicity gradients include galactic winds, which eject metals out of galaxies \citep[e.g.,][]{Hayward2017, Pandya2021}, and radial advection, which transports pristine or metal-rich gas across the disk in bulk flows \citep[e.g.,][]{Sharda2021}. Together, these processes highlight the dynamic and interconnected nature of gas mixing and its role in shaping metallicity profiles.

\begin{figure*}[htb!]
    \centering
    \includegraphics[width=0.49\linewidth]{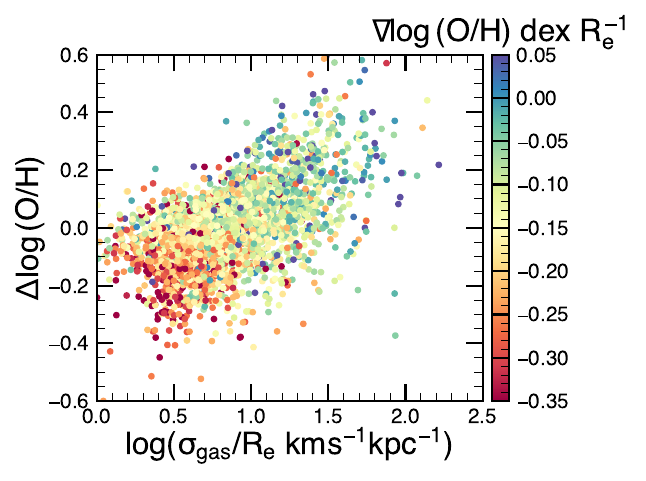}
    \includegraphics[width=0.49\linewidth]{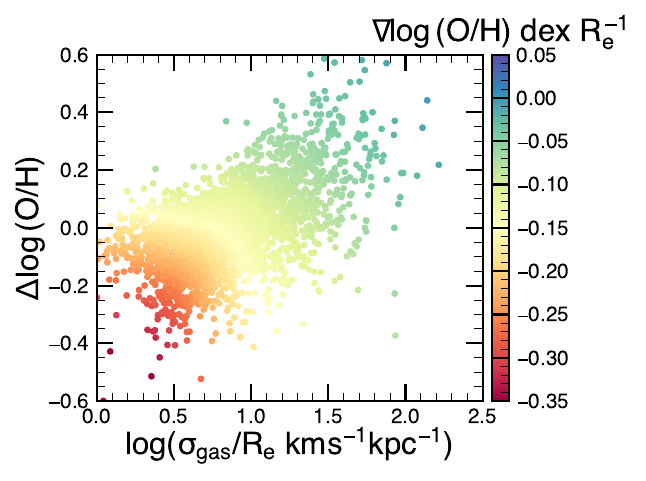}
    \includegraphics[width=1.0\linewidth]{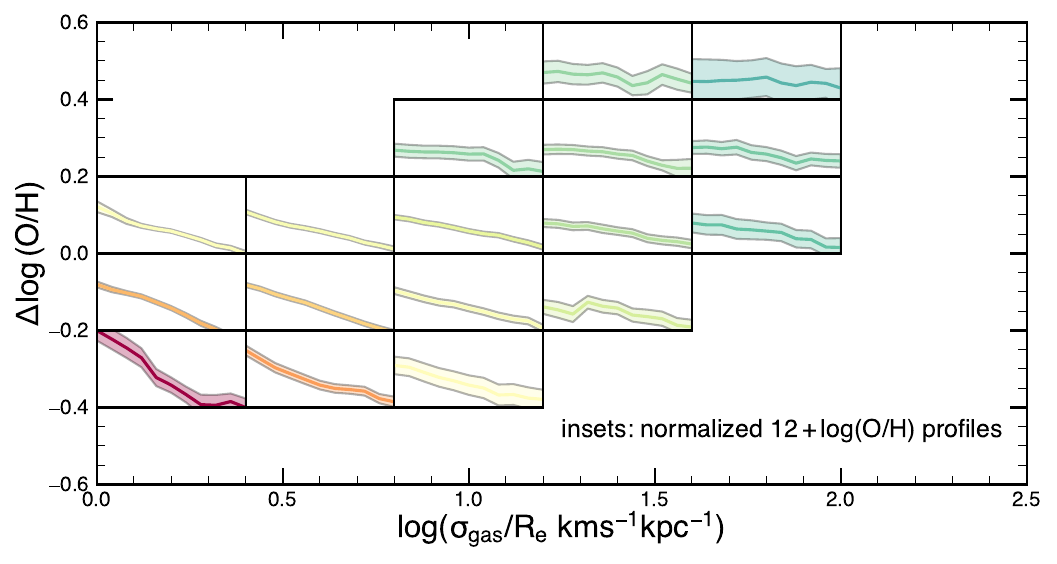}
    \caption{Metallicity gradient on the disk as functions of metallicity excess at the effective radius and local gas turbulence. The top panels show the scatter plot of each galaxy and the corresponding smooth results. The dots and colors are similar to the top panels of Figure \ref{MZR}. The bottom panel depicts the median gas-phase metallicity profiles, each normalized to arbitrary values, within bins of 0.4 dex in $\log(\sigma_\mathrm{gas}/R_\mathrm{e})$ and 0.2 dex in $\Delta \log(\mathrm{O/H})$. Each inset maintains a consistent y-axis range, with the x-axis spanning from 0.5 to 1.5 $R_\mathrm{e}$. The color of each profile corresponds to $\nabla \log(\mathrm{O/H})$, denoted in the colorbar with shaded areas indicating the uncertainties of the median values.}
    \label{Zgasexc_sigmagasRe}
\end{figure*}

Motivated by these processes, we define $\sigma_{\mathrm{gas}}/R_{\mathrm{e}}$ as a proxy for local gas turbulence within the galaxy disk. This quantity encapsulates the strength of gas turbulence at galactic scale for individual galaxies, which serves as an imprint of various physical processes. Gas turbulence tends to flatten metallicity gradients in star-forming galaxies through various physical processes that drive gas mixing and enrichment. However, distinguishing between these processes is challenging without numerous physical assumptions. In Figure \ref{Zgasexc_sigmagasRe}, we plot the metallicity gradient as a function of metallicity excess and $\sigma_{\mathrm{gas}}/R_{\mathrm{e}}$. We find that $\nabla \log(\mathrm{O/H})$ correlates strongly with both $\Delta \log(\mathrm{O/H})$ and $\log(\sigma_{\mathrm{gas}}/R_{\mathrm{e}})$, consistent with the diffusion effect described by \citet{Wang2022b}, where larger turbulent velocities lead to more flattened metallicity profiles. These trends are clearly depicted through the median metallicity profiles across different ranges of $\sigma_{\mathrm{gas}}/R_{\mathrm{e}}$ and $\Delta \log(\mathrm{O/H})$, as presented in the bottom panel of Figure \ref{Zgasexc_sigmagasRe}. We further quantify the combined effect of $\sigma_{\mathrm{gas}}/R_{\mathrm{e}}$ and $\Delta \log(\mathrm{O/H})$ in shaping metallicity gradients in Appendix \ref{fXY} and consider alternative combinations of physical properties in predicting metallicity gradient in Appendix \ref{AlternativeCom}. The residual scatters in Figure \ref{scatter-pars} within Appendix \ref{AlternativeCom} further demonstrate that $\Delta \log(\mathrm{O/H})$ is strongly related to the metallicity gradient, with $\sigma_{\mathrm{gas}}$ and $R_{\mathrm{e}}$ being comparably informative on the metallicity gradient.

These findings suggest that the gas-phase metallicity gradient results from competing effects of coplanar gas inflow and local gas turbulence on the disk. These two physical quantities co-evolve through the interplay of processes, such as gas inflow, star formation, metal enrichment, gas outflows, and gas mixing. Consequently, there is also a positive correlation between metallicity excess and $\sigma_{\mathrm{gas}}/R_{\mathrm{e}}$, as shown in Figure \ref{Zgasexc_sigmagasRe}. This relationship underscores the complex interactions between turbulence and chemical enrichment in galaxy disks.

\subsection{Testing feature importance with Random Forest}

Previous studies have inferred the roles of various physical processes in shaping metallicity gradients by examining correlations with simple, reliable galaxy properties. Observational evidence indicates that metallicity gradients correlate with a range of galaxy characteristics, including stellar mass \citep[e.g.,][]{Belfiore2017, Mingozzi2020, Schaefer2020, Franchetto2021}, galaxy size \citep[e.g.,][]{Carton2018, Boardman2021, Boardman2022, Boardman2023, Lin2024}, gas fraction \citep[e.g.,][]{Franchetto2021}, bulge-to-total (B/T) ratio \citep[e.g.,][]{Moran2012}, and environmental factors \citep[e.g.,][]{Lian2019, Franchetto2021}. Identifying a concise set of physical quantities that encapsulate or reflect the impact of these processes is critical to understanding the factors driving metallicity gradients. 

\begin{figure}[htb!]
    \centering
    \includegraphics[width=1\linewidth]{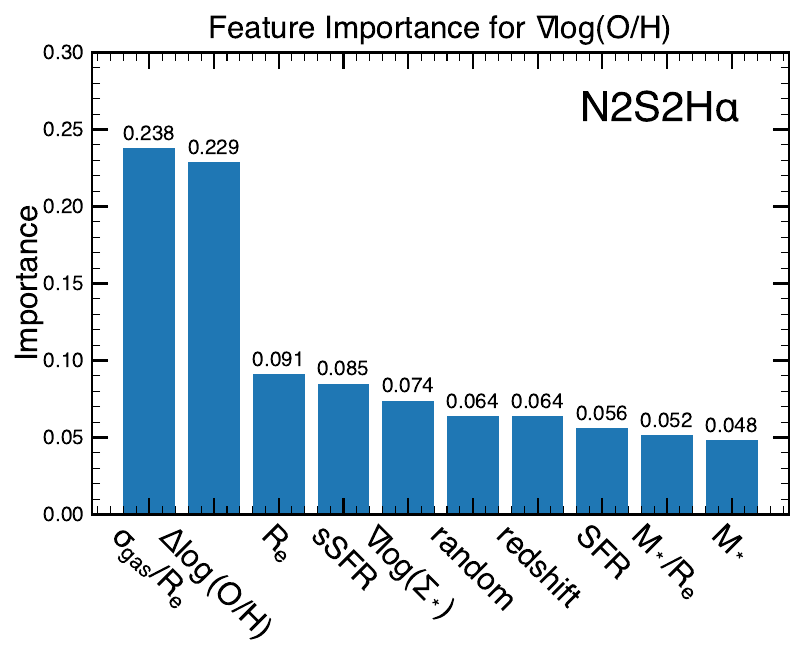}
    \caption{Relative importance of galaxy physical parameters (labeled on the x-axis) employed for regressing the metallicity gradient in Random Forest models. The metallicity calibration is marked in the top-right corner.}
    \label{RF}
\end{figure}

To investigate whether metallicity excess and local gas turbulence sufficiently describe metallicity gradients, we employ a Random Forest model. This machine-learning approach allows us to assess the contributions of different galaxy properties to the metallicity gradient. In this context, relative feature importance quantifies the contribution of input features to the model's predictions. Using the \texttt{scikit-learn} implementation, feature importance is ranked based on the Gini importance \citep{Pedregosa2011}. Figure \ref{RF} indicates that metallicity excess and $\sigma_{\mathrm{gas}}/R_{\mathrm{e}}$ dominate the feature importance rankings. These two quantities significantly outperform other commonly used galaxy properties, such as stellar mass, effective radius, star formation rate, 
and effective radius, in predicting metallicity gradients. While the $\nabla\log\Sigma_*$ plays a relatively minor role, consistent with the trends in Figure \ref{Zgasexc_SDslope_ccZgasslope}. We have also checked to perform an alternative Random Forest model with independent parameters ($\sigma_{\mathrm{gas}}$, $M_*$, $R_\mathrm{e}$, $\Delta \log\mathrm{(O/H)}$, SFR, redshift, $\nabla\log\Sigma_*$, and a random variable). The analysis consistently identifies $\Delta \log\mathrm{(O/H)}$, $\sigma_{\mathrm{gas}}$, and $R_\mathrm{e}$ as the top three most important features, which aligns with the results in Figure \ref{RF}. Additionally, we have tested alternative metallicity indicators and calibrations to calculate both metallicity gradients and metallicity values, finding consistent results (see Appendix \ref{AlternativeZ}). 

This consistency supports the robustness of our findings that metallicity excess and $\sigma_{\mathrm{gas}}/R_{\mathrm{e}}$ as predictive features reinforce the role of coplanar gas inflow in disk formation and growth. These results also support the hypothesis that coplanar gas inflow plays a dominant role in driving the disk evolution of star-forming galaxies.

\section{Summary} \label{sec:summary}

In this work, we analyze star-forming galaxies from the MaNGA sample to identify the key parameters influencing the radial gradients of gas-phase metallicity. Our goal is to determine whether disk formation is primarily driven by coplanar gas inflow or by the independent evolution of annuli within the disk. The main findings are summarized below:

(1) The metallicity gradient on the disk strongly correlates with the local gas-phase metallicity for star-forming galaxies at a given stellar mass. Specifically, $\nabla \log(\mathrm{O/H})$ is steeper and more negative for metal-poorer disks (Figure \ref{MZR}). This trend supports the scenario of coplanar gas inflow, where gas is progressively enriched by in situ star formation as it flows inward through the disk.

(2) The radial gradient of stellar mass surface density shows a fairly weak correlation with $\nabla \log(\mathrm{O/H})$ (Figure \ref{Zgasexc_SDslope_ccZgasslope}). This weak correlation suggests that gas inflow, star formation, and metal enrichment interact in a coupled manner within each annulus of the disk, rather than evolving independently.

(3) Competing effects of coplanar gas inflow and local gas turbulence (indicated by $\sigma_{\mathrm{gas}}/R_{\mathrm{e}}$) on the disk can well describe the metallicity gradient (Figure \ref{Zgasexc_sigmagasRe}). The primary role of $\Delta \log(\mathrm{O/H})$ in shaping $\nabla \log(\mathrm{O/H})$ is confirmed by the feature importance in the Random Forest model (Figure \ref{RF}).

Our results serve as indirect observational imprints for gas inflow and support the dominating role of coplanar gas inflow in driving disk evolution. However, the dominance of coplanar inflow may not extend to high-redshift galaxies. High-redshift galaxies often exhibit irregular morphologies and lack azimuthally symmetric chemical distributions due to frequent mergers and asymmetric infall of low-mass gas \citep[e.g.,][]{Bellardini2021}. Such conditions make traditional radial gradients less effective for quantifying metal distributions. Furthermore, high-redshift galaxies are subject to more frequent mergers and elevated turbulence driven by intense star formation, which may hinder the formation of well-defined disk structures observed in the local Universe \citep[e.g.,][]{Rodriguez-Gomez2015, Wisnioski2015}. Additionally, small-scale structures, such as spiral arms and bars, could introduce significant variations in radial metallicity profiles.

Even though, recent observations enabled by the James Webb Space Telescope (JWST) have revealed flatter metallicity gradients at high redshift. These findings, derived from rest-frame optical emission lines, suggest that elevated gas velocity dispersion and enhanced turbulence may play a dominant role in shaping metallicity distributions at early epochs \citep[e.g.,][]{Wang2020, Wang2022a, RodriguezDelPino2024, Venturi2024, Ju2024}. Future studies should focus on developing alternative tracers for spatially resolved metallicity to complement simulation predictions \citep[e.g.,][]{Hemler2021, Tissera2022, Sun2024} and facilitate meaningful comparisons with high-redshift observations. This approach may open new parameter spaces for understanding and testing disk formation at early cosmic times.

\section{Acknowledgments}.

\begin{acknowledgments}
The authors thank the anonymous referee for their helpful comments that improved the quality of this paper. EW thanks support of the National Science Foundation of China (Nos. 12473008) and the Start-up Fund of the University of Science and Technology of China (No. KY2030000200). The authors gratefully acknowledge the support of Cyrus Chun Ying Tang Foundations. YP acknowledges support from the National Science Foundation of China (NSFC) grant Nos. 12125301, 12192220 and 12192222, and support from the New Cornerstone Science Foundation through the XPLORER PRIZE. XW is supported by the National Natural Science Foundation of China (grant 12373009), the CAS Project for Young Scientists in Basic Research Grant No. YSBR-062, the Fundamental Research Funds for the Central Universities, the Xiaomi Young Talents Program, and the science research grant from the China Manned Space Project. 

Funding for the Sloan Digital Sky Survey IV has been provided by the Alfred P. Sloan Foundation, the U.S. Department of Energy Office of Science, and the Participating Institutions. SDSS-IV acknowledges support and resources from the Center for High-Performance Computing at the University of Utah. The SDSS web site is www.sdss.org.

SDSS-IV is managed by the Astrophysical Research Consortium for the Participating Institutions of the SDSS Collaboration including the Brazilian Participation Group, the Carnegie Institution for Science, Carnegie Mellon University, the Chilean Participation Group, the French Participation Group, HarvardSmithsonian Center for Astrophysics, Instituto de Astrofísica de Canarias, The Johns Hopkins University, Kavli Institute for the Physics and Mathematics of the Universe (IPMU) / University of Tokyo, the Korean Participation Group, Lawrence Berkeley National Laboratory, Leibniz Institut für Astrophysik Potsdam (AIP), Max-Planck-Institut für Astronomie (MPIA Heidelberg), Max-Planck-Institut für Astrophysik (MPA Garching), Max-Planck-Institut für Extraterrestrische Physik (MPE), National Astronomical Observatories of China, New Mexico State University, New York University, University of Notre Dame, Observatário Nacional / MCTI, The Ohio State University, Pennsylvania State University, Shanghai Astronomical Observatory, United Kingdom Participation Group, Universidad Nacional Autónoma de México, University of Arizona, University of Colorado Boulder, University of Oxford, University of Portsmouth, University of Utah, University of Virginia, University of Washington, University of Wisconsin, Vanderbilt University, and Yale University.
\end{acknowledgments}

\vspace{5mm}

\appendix

\section{Linear combination of metallicity excess and local gas turbulence}\label{fXY}

\begin{figure}[htb!]
    \centering
    \includegraphics[width=0.49\linewidth]{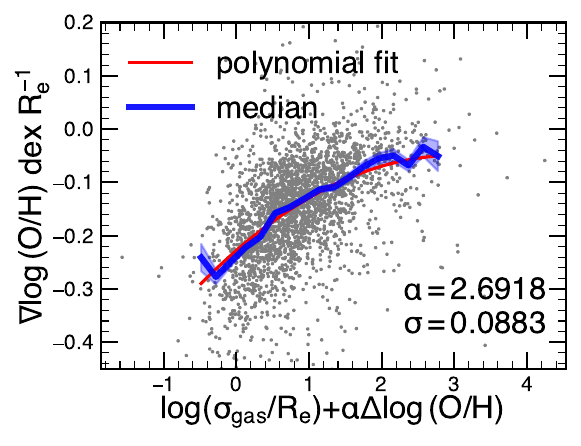}
    \caption{Metallicity gradient on the disk as a function of the linear combination of local turbulence and metallicity excess at the effective radius. Each dot represents an individual galaxy in our sample, with $\alpha \sim 2.69$ minimizing the residual scatter ($\sigma \sim 0.088$) of the relation. The blue curve indicates the median values with shaded area as uncertainties and the red curve represents the 2nd polynomial fit of the data.}
    \label{Z_fXY}
\end{figure}

To further quantify the combined effect of local gas turbulence and coplanar inflow in shaping metallicity gradients, we introduce a new parameter, $\mu_\alpha$, defined as a linear combination of local gas turbulence and metallicity excess: $\mu_\alpha=\log(\sigma_\mathrm{gas}/R_\mathrm{e})+\alpha(\Delta \log(\mathrm{O/H})),$ where $\alpha$ is a free parameter. The optimal value of $\alpha$ minimizes the scatter in the median metallicity gradient relative to this parameter. We find that $\alpha \sim 2.69$ achieves the minimum dispersion. The relationship between the metallicity gradient and $\mu_\alpha$ is shown in Figure \ref{Z_fXY}, where the improved correlation highlights the combined influence of local gas turbulence and metallicity excess on the metallicity gradient.

\section{Alternative combinations of physical properties in predicting metallicity gradient} 
\label{AlternativeCom}

\begin{figure*}[htb!]
    \centering
    \includegraphics[width=1.0\linewidth]{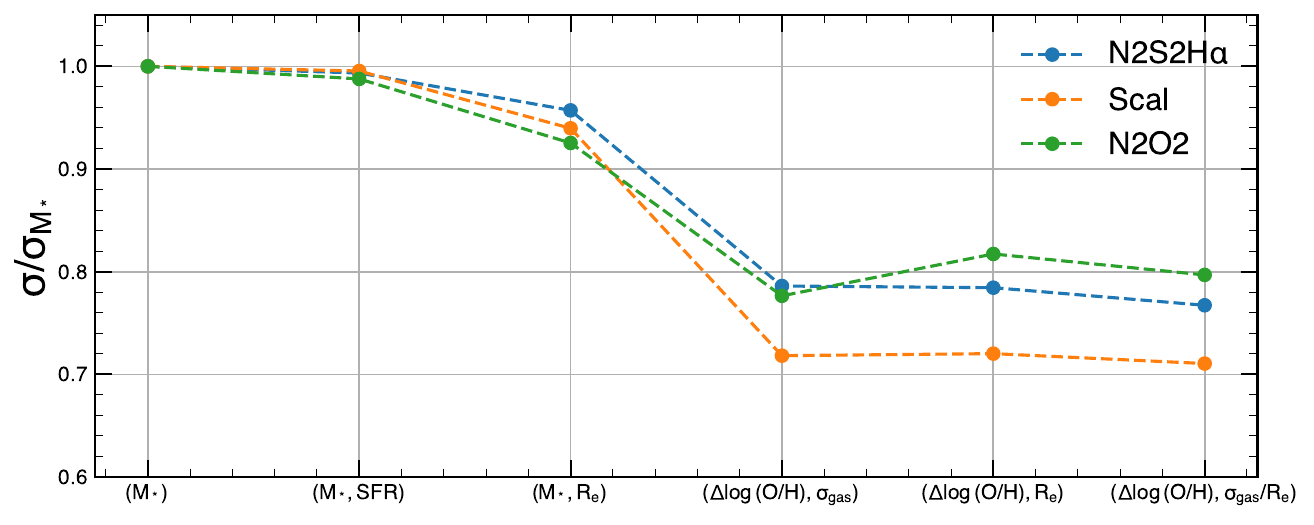}
    \caption{The residual scatters to the fitting curves of metallicity gradient versus $\mu_\alpha=Y+\alpha(X)$ with various combinations of physical parameters $X$ and $Y$ ($Y$, $X$ in logarithmic scale) labeled on the x-axis. The residual scatters are normalized to $\sigma_{M_*}$ ($X=0$, $Y=\log(M_*/M_\odot)$). Different colors indicate the curves for three different gas-phase metallicity estimators or calibrations.}
    \label{scatter-pars}
\end{figure*}

\begin{figure*}[htb!]
    \centering
    \includegraphics[width=0.49\linewidth]{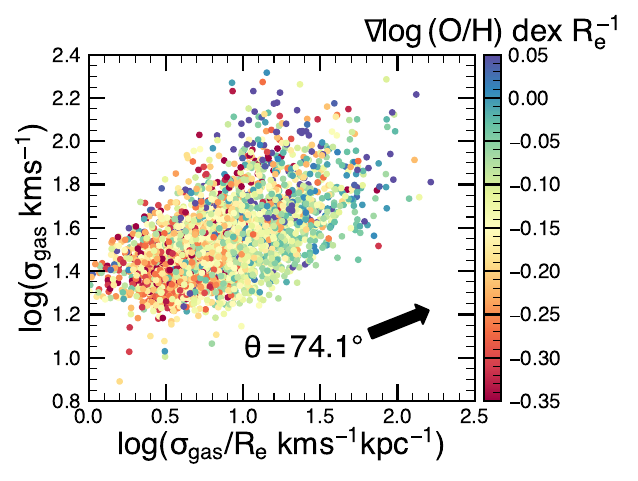}
    \includegraphics[width=0.49\linewidth]{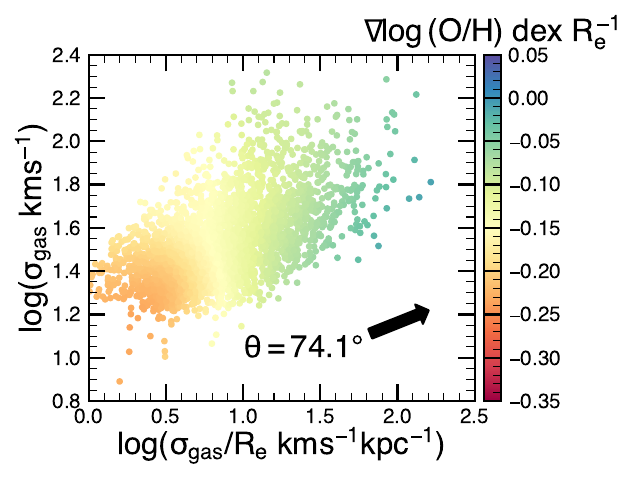}
    \caption{Metallicity gradient as functions of $\sigma_\mathrm{gas}/R_\mathrm{e}$ and $\sigma_\mathrm{gas}$. The left and right
    panels show the scatter plot and corresponding smooth results, respectively. The dots and colors are similar to the top panels of Figure \ref{MZR}. The color gradients, indicated by arrows, represent the directions of increasing $\nabla \log\mathrm{(O/H)}$.}
    \label{sigmaRe_sigma}
\end{figure*}

Similar to the analysis in \citet{Ma2024}, we quantify the combined effect in determining the metallicity gradient considering alternative combinations of physical properties in predicting metallicity gradient. As shown in Figure \ref{scatter-pars}, the scatter in Figure \ref{fXY} ($\Delta \log\mathrm{(O/H)}$, $\log(\sigma_\mathrm{gas}/R_\mathrm{e})$) is lower than that observed for the combinations of ($M_*$, $R_\mathrm{e}$) or ($M_*$, SFR), and is comparable to or slightly lower than the combinations of ($\Delta \log\mathrm{(O/H)}$, $\sigma_\mathrm{gas}$) or ($\Delta \log\mathrm{(O/H)}$, $R_\mathrm{e}$). This result underscores the important roles of $\Delta \log\mathrm{(O/H)}$ and $\log(\sigma_\mathrm{gas}/R_\mathrm{e})$ in shaping metallicity gradient, consistent with our Random Forest analysis. Note that although the scatter drops slightly from combinations of ($\Delta \log\mathrm{(O/H)}$, $\sigma_\mathrm{gas}$) or ($\Delta \log\mathrm{(O/H)}$, $R_\mathrm{e}$) to ($\Delta \log\mathrm{(O/H)}$, $\log(\sigma_\mathrm{gas}/R_\mathrm{e})$), we still consider $\sigma_\mathrm{gas}/R_\mathrm{e}$ as an important feature because it integrates both the degree of turbulence and the size scale, providing richer physical significance. For instance, the same level of turbulence distributed over a large scale would be diluted, whereas the same turbulence concentrated on a small scale could have a more significant impact on star formation and gas flows. As illustrated in Figure \ref{sigmaRe_sigma}, $\sigma_\mathrm{gas}/R_\mathrm{e}$ exhibits a significantly tighter with metallicity gradients compared to $\sigma_\mathrm{gas}$ alone. This trend further underscores the importance of $\sigma_\mathrm{gas}/R_\mathrm{e}$ as a parameter of interest. Additionally, the unit of $\sigma_\mathrm{gas}/R_\mathrm{e}$ is the reciprocal of time, which can be compared to the star formation timescale or gas depletion timescale. We plan to explore these connections in future work.

\section{Feature importance from other metallicity calibrations} \label{AlternativeZ}

\begin{figure*}[htb!]
    \centering
    \includegraphics[width=0.48\linewidth]{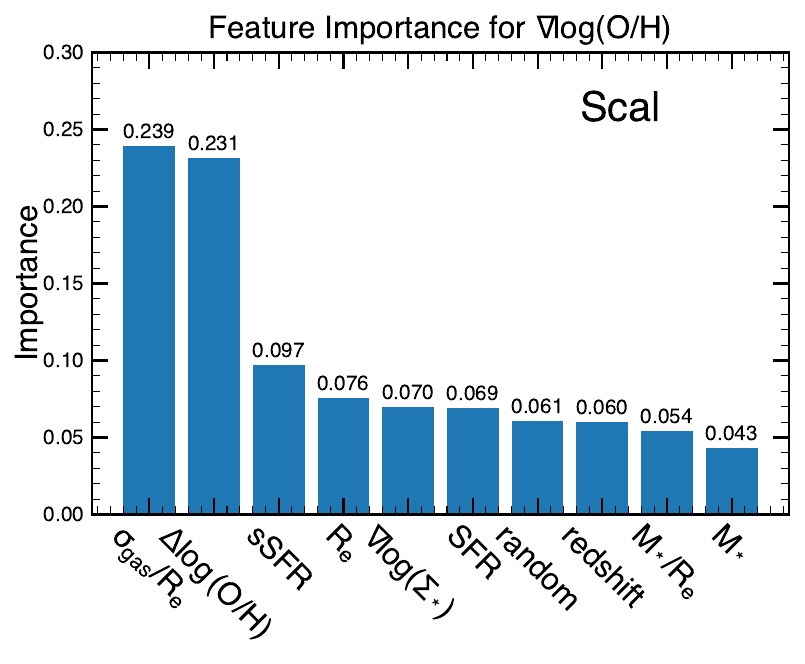}
    \includegraphics[width=0.48\linewidth]{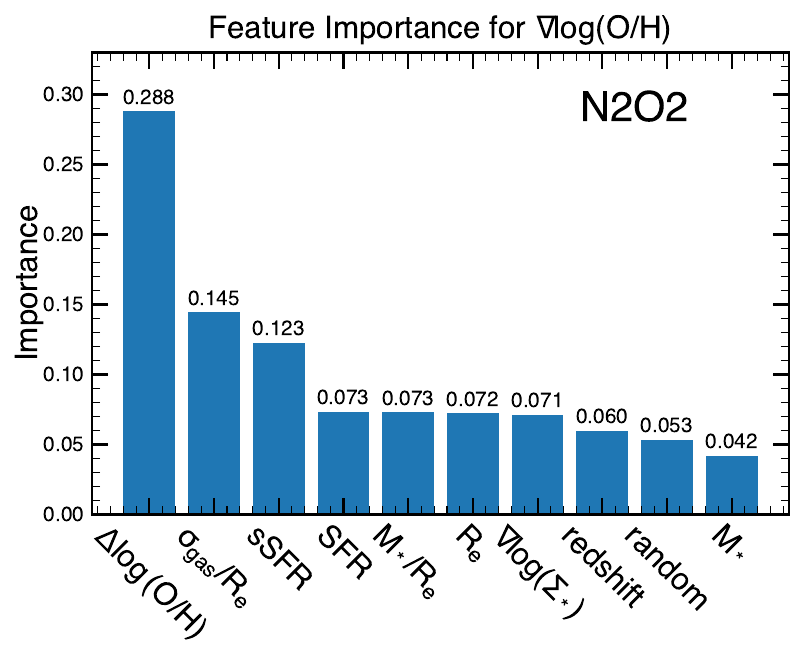}
    \caption{Relative importance of galaxy physical parameters (labeled on the x-axis) employed for regressing the metallicity gradient in Random Forest models. $\nabla \log(\mathrm{O/H})$ and $\Delta\log(\mathrm{O/H})$ are calculated by alternative estimators or calibrations marked in the top-right corner of each panel.}
    \label{RF-A}
\end{figure*}

Similar to Figure \ref{RF}, we test two alternative common-used metallicity indicators to calculate both metallicity gradients and metallicity values and plot the feature importance in regressing metallicity gradient. As shown in Figure \ref{RF-A}, results based on S-calibration \citep[\texttt{Scal}, ][left panel]{Pilyugin2016} and \texttt{N2O2} diagnostic \citep[][right panel]{Dopita2013, Zhang2017} consistently reveal that metallicity excess and local gas turbulence are two primary quantities in shaping $\nabla\log\mathrm{(O/H)}$. These results align with the hypothesis that coplanar gas inflow dominates the disk evolution in star-forming galaxies.

\bibliography{sample631}{}
\bibliographystyle{aasjournal}

\end{document}